\documentclass[prb,twocolumn,showpacs]{revtex4}

\usepackage{graphicx,epsf}% Include figure files
\usepackage{bm}% bold math

\begin{document}

\title{Coulomb blockade and non-Fermi-liquid behavior in quantum dots}
\smallskip

\author{Frithjof B. Anders,$^1$ Eran Lebanon,$^2$ and Avraham Schiller$^2$}
\affiliation{$^1$Department of Physics, Universit\"at Bremen,
                  P.O. Box 330 440, D-28334 Bremen, Germany\\
             $^2$Racah Institute of Physics, The Hebrew University,
                  Jerusalem 91904, Israel}

\begin{abstract}
The non-Fermi-liquid properties of an ultrasmall quantum dot
coupled to a lead and to a quantum box are investigated. Tuning
the ratio of the tunneling amplitudes to the lead and box, we
find a line of two-channel Kondo fixed points for arbitrary
Coulomb repulsion on the dot, governing the transition between
two distinct Fermi-liquid regimes. The Fermi liquids are
characterized by different values of the conductance. For an
asymmetric dot, spin and charge degrees of freedom are entangled:
a continuous transition from a spin to a charge two-channel Kondo
effect evolves. The crossover temperature to the two-channel Kondo
effect is greatly enhanced away from the local-moment regime,
making this exotic effect accessible in realistic quantum-dot devices. 
\end{abstract}

\maketitle

% General introduction -- quantum criticality and the 2CKE
Strongly correlated electron systems display non-Fermi-liquid
properties in the vicinity of a zero-temperature phase
transition.~\cite{Hertz76,Millis93} In lattice systems, the
nature of these so-called quantum critical points is still
not well understood. While theoretical descriptions typically
start from well-defined quasiparticle excitation modes, the
non-Fermi-liquid behavior is interaction driven, and arises
from persisting quantum-mechanical fluctuations between these
modes.~\cite{Hertz76,Millis93} The two-channel Kondo effect
(2CKE) is a prototype for such a quantum critical point in
a quantum
impurity system. It occurs when a spin-$\frac{1}{2}$ local
moment is coupled antiferromagnetically with equal strength
to two independent conduction-electron channels that
overscreen the moment.~\cite{CZ98} Its fixed point governs
the transition between two distinct Fermi liquids, which are
adiabatically connected at finite temperature. While certain
ballistic metallic point contacts, two-level systems, and heavy
fermion alloys have been argued to display the 2CKE,~\cite{CZ98}
a conclusive experimental observation of this elusive effect
remains lacking.

% More specific introduction -- nano-devices and the 2CKE
Quantum-dot devices have become an important tool for
investigating fundamental questions such as the 2CKE, since
they allow for detailed sample engineering and direct control
of the microscopic model parameters. Here we investigate
realizations of the 2CKE in a double-dot device, comprised
of a quantum box (i.e., a large dot) indirectly coupled to
a lead via an ultrasmall quantum dot.~\cite{LSA03} In the
Coulomb-blockade valleys, charge fluctuations are suppressed
on the quantum box. This blocks the interlead exchange
coupling at temperatures below the charging energy of the box,
thereby generating two independent screening channels for
the local moment formed on the small dot.~\cite{OGG03} A spin
2CKE then develops on the dot if the effective spin-exchange
couplings to the lead and box are tuned to be equal.~\cite{OGG03}
So far, only the local-moment regime was considered within
this scenario.~\cite{OGG03,PBGvD03,FR03}

% goal and first summary
In this paper, we explore different regimes of the lead--dot--box
device, where charge is not quantized on the quantum dots.
We find the following:
(i) a line of two-channel fixed points as a function of the gate
    voltages in the device, extending to the mixed-valent regime
    of the ultrasmall dot and away from the Coulomb-blockade
    valleys of the quantum box;
(ii) a continuous transition from a spin 2CKE to a charge 2CKE
     for an asymmetric dot;
(iii) an intriguing entanglement of spin and charge within the
      2CKE that develops in the experimentally relevant case of
      particle-hole asymmetry, reflected in a simultaneous
      $\ln(T)$ divergence of the magnetic susceptibility of
      the dot and the charge capacitance of the box;
(iv) an abrupt jump in the $T = 0$ conductance across the
     two-channel line. Here the Fermi liquids on either side of
     the critical line are characterized by distinct values of
     the $T = 0$ conductance.

% model
The setting we consider consists of an ultrasmall quantum dot,
modeled by a single energy level $\epsilon_d$ and an onsite
repulsion $U$, embedded between a metallic lead and a quantum
box. The quantum box is characterized by a finite charging
energy, $E_C$, and by a dense set of single-particle levels,
which we take to be continuous. Denoting the creation of an
electron with spin projection $\sigma$ on the dot by
$d_{\sigma}^{\dagger}$, the corresponding Hamiltonian reads
\begin{eqnarray}
&& {\cal H} = \sum_{\alpha=L,B} \sum_{k,\sigma}
                     \epsilon_{\alpha k} 
                     c^{\dagger}_{\alpha k\sigma}c_{\alpha k\sigma}
           + \; E_C \left ( \hat{n}_B - N_B \right )^2
\label{initial_Hamiltonian} \\
&& +\
      \epsilon_d \sum_{\sigma} d^{\dagger}_{\sigma} d_{\sigma}
           + U \hat{n}_{d \uparrow} \hat{n}_{d \downarrow}
      + \sum_{\alpha, k, \sigma} t_{\alpha}
      \left\{ 
            c^{\dagger}_{\alpha k\sigma}d_{\sigma} + {\rm H.c.}
      \right\} ,
\nonumber 
\end{eqnarray}
where $c^{\dagger}_{L k \sigma}$ ($c^{\dagger}_{B k \sigma}$) creates 
a lead (box) electron with momentum $k$ and spin projection $\sigma$,
$t_L$ ($t_B$) is the tunneling matrix element between the quantum dot
and the lead (box), $\epsilon_{L k}$ ($\epsilon_{B k}$) are the
single-particle levels in the lead (box), and $\hat{n}_{d\sigma}$
equals $d^{\dagger}_{\sigma}d_{\sigma}$. The excess number of
electrons inside the box, $\hat{n}_B = \sum_{k\sigma}
[c^{\dagger}_{Bk\sigma}c_{Bk\sigma}-\theta(-\epsilon_{Bk})]$,
is controlled by the dimensionless gate voltage, $N_B$.

% mapping
We treat the Hamiltonian of Eq.~(\ref{initial_Hamiltonian})
using a recent adaptation of Wilson's numerical
renormalization-group (NRG) method~\cite{Wilson75} to the
Coulomb blockade.~\cite{LSA03_FR} Introducing three collective
charge operators for the box,~\cite{Schoeller_etal}
$\hat{N}\! =\! \sum_{m = -\infty}^{\infty}
m |m \rangle\langle m |$
and
$\hat{N}^{\pm}\! =\! \sum_{m = -\infty}^{\infty}
|m \pm 1 \rangle\langle m |$,
we replace the tunneling term between the dot and the box in 
Eq.~(\ref{initial_Hamiltonian}) with
$\sum_{k,\sigma} t_{B} \left \{ \hat{N}^{+}
c^{\dagger}_{B k \sigma} d_{\sigma} + {\rm H.c.}\right \}$,
while the charging-energy term is converted to
$E_C (\hat{N} - N_B)^2$. Since the dynamics of $\hat{N}$ is
insensitive to the precise number of conduction electrons in
the bands, one can relax the constraint $\hat{N} = \hat{n}_B$,
and regard $\hat{N}$ as an independent degree of
freedom.~\cite{LSA03_FR} The resulting Hamiltonian describes
then two noninteracting conduction bands coupled to a
complex impurity, composed of both $\hat{N}$ and the dot
degrees of freedom. Taking the two conduction baths to
have a common rectangular density of states,
$\rho(\epsilon) = \rho \theta ( D - |\epsilon|)$, we solve
the resulting model using the NRG. Here the number $N_s$
of NRG states retained sets a lower bound on the ratio
$E_C/D$ one can treat. Throughout the paper we set the NRG
discretization parameter~\cite{Wilson75} $\Lambda$ equal to
$2.8$, while $N_s = 2000$. 

\begin{figure}
\centerline{
\includegraphics[width=75mm]{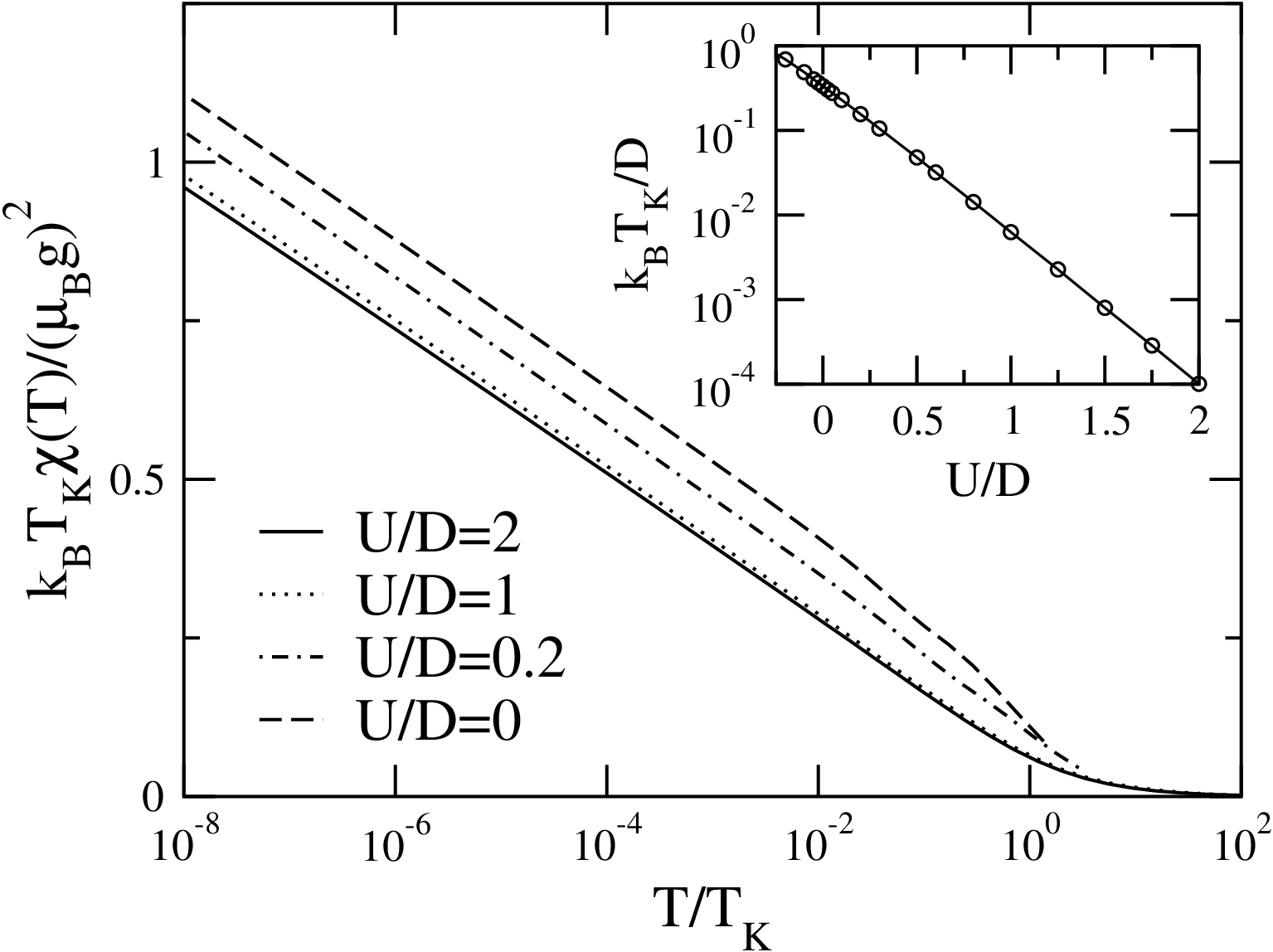}
}\vspace{-5pt}
\caption{The magnetic susceptibility of the dot vs $T$, for
         $\Gamma_L = E_C = 0.1D$, $N_B = 0$, and different values
         of $U = -2\epsilon_d$. Here $\Gamma_B$ is tuned for each
         value of $U$ to the two-channel point $\Gamma_B^{\rm 2CK}$.
         The Kondo temperature $T_K$ is defined in
         Eq.~(\ref{Tk-def}). For all $U$, ranging
         from the local-moment regime $U \gg \Gamma_L$, to the strongly
         mixed-valent regime $U = 0$, to negative $U$ (not shown),
         there exists a two-channel point $\Gamma_B^{\rm 2CK}$
         where $\chi(T)$ diverges logarithmically as $T \to 0$, and
         the finite-size spectrum converges to the conventional
         two-channel fixed point. Inset: $T_K$ vs $U$. For
	 $U \gg \Gamma_L, E_C$ (local-moment regime), $T_K$ decreases
	 exponentially with $U$. For $U < \Gamma_L$ (mixed-valent
	 regime), it exceeds $E_C$.}
\label{fig:fig1}
\end{figure}

We begin with a symmetric dot, $U = - 2\epsilon_d$, and with a
box tuned to the middle of a Coulomb-blockade valley, $N_B = 0$.
Fixing $E_C$ and the hybridization strength to the lead,
$\Gamma_L = \pi \rho t_L^2$, we varied the hybridization to the
box, $\Gamma_B = \pi \rho t_B^2$, in search of a 2CKE. Our results
for $E_C = \Gamma_L = 0.1D$ are summarized in Fig.~\ref{fig:fig1}.
Quite surprisingly, we find a two-channel Kondo point
$\Gamma_B^{\rm 2CK}$ for all values of $U$, ranging from the
local-moment regime $U \gg \Gamma_L$, to the strongly mixed-valent
regime $U \approx 0$, to the unrealistic regime of $U < 0$. The
development of a 2CKE is reflected in a logarithmic divergence
of the dot susceptibility~\cite{comment_on_susceptibility} as
$T \to 0$ (see Fig.~\ref{fig:fig1}), and in a finite-size
spectrum that converges to the conventional two-channel fixed
point (i.e., identical energies, degeneracies, and quantum
numbers).

The emergence of a spin 2CKE for $U \gg \Gamma_L$ proves the
scenario of Ref.~\onlinecite{OGG03}. Specifically, the
associated Kondo temperature decays exponentially with $U$ for
$U \gg \Gamma_L, E_C$ (inset of Fig.~\ref{fig:fig1}), while the
ratio $\Gamma_B^{\rm 2CK}/\Gamma_L$ approaches the asymptotic
value $1 + 2E_C/U$ (not shown). Here and throughout the
paper we define the Kondo temperature $T_K$ according to the
Bethe ansatz expression for the slope of the $\ln(T)$ diverging
term in the susceptibility of the two-channel Kondo
model,~\cite{SS89}
\begin{equation}
\chi(T) \sim \frac{(\mu_B g)^2}{20 k_B T_K}\ln(T_K/T) .
\label{Tk-def}
\end{equation}
We emphasize, however, that our NRG calculations go well beyond the
Schrieffer-Wolff transformation used in Ref.~\onlinecite{OGG03},
confirming that no relevant perturbations are generated at higher
orders in the tunneling amplitudes. 

Contrary to the local-moment regime, the development of a spin
2CKE in the mixed-valent regime (let alone for $U < 0$) is an
unexpected feature, with no apparent spin degree of freedom to
be overscreened. Moreover, $T_K$ is significantly enhanced for
$U \approx 0$, exceeding $E_C$ in Fig.~\ref{fig:fig1}. This
behavior is reminiscent of the two-channel Anderson model,
where the 2CKE likewise persists into the mixed-valent
regime.~\cite{2ch_Anderson} Particularly intriguing is the
limit of a noninteracting dot, which reduces~\cite{LSA03} to
the familiar problem of a quantum box connected to a lead by
single-mode tunneling. Although a charge 2CKE was predicted
for the latter setting at the degeneracy points of the Coulomb
blockade,~\cite{Matveev91} no spin 2CKE was previously
anticipated.

To understand the 2CKE for $U = 0$, we revisit the problem
of a quantum box connected to a lead by a nearly fully
transmitting single-mode point contact. Following
Refs.~\onlinecite{Flensberg93} and \onlinecite{Matveev95},
we model this system by a one-dimensional (1D) geometry,
where $x < 0$ ($x > 0$)
represents the lead (box), and $x = 0$ corresponds to the
noninteracting dot. The deviation from perfect transmission
is modeled~\cite{Flensberg93,Matveev95} by weak
backscattering at $x = 0$. Extending the
bosonization treatment of Ref.~\onlinecite{Matveev95} to
a finite local magnetic field acting on the dot, and
carefully accounting for an underlying symmetry of the
Hamiltonian,~\cite{forthcoming} we obtain the following
linear magnetic susceptibility for $k_B T \ll E_C$:
\begin{equation}
\chi(T) = \chi_0 \left( \frac{\mu_B g}{\hbar v_F} \right)^2\!
          \left [
           \ln \! \left ( \frac{D_{\rm eff}}{2\pi k_B T} \right )\!
           - \psi \! \left( \frac{1}{2}
                            + \frac{\Gamma}{2\pi k_B T}
                     \right)
          \right ] ,
\label{chi_Matveev}
\end{equation}
where $\chi_0 = \Gamma/(4 \pi R)$ and
$\Gamma = R (8\gamma E_C/\pi^2) \cos^2(\pi N_B)$.
Here, $R$ is the reflectance, $v_F$ is the Fermi velocity,
$\gamma$ equals $e^C$ with $C \approx 0.5772$, $\psi(x)$ is
the digamma function, and $D_{\rm eff}$ is an effective cutoff
of the order of $E_C$.

Equation~(\ref{chi_Matveev}) features a
logarithmic temperature dependence down to $k_B T \sim \Gamma$.
Hence, $\chi(T)$ diverges logarithmically for perfect transmission,
when $\Gamma \propto R$ vanishes. In fact, $\chi(T)$ diverges
logarithmically at perfect transmission for all gate voltages,
except for half-integer values of $N_B$ where $\chi_0$ vanishes. In
particular, Eq.~(\ref{chi_Matveev}) predicts
$T_K(N_B) \propto 1/\cos^{2}(\pi N_B)$.

\begin{figure}
\centerline{
\includegraphics[width=70mm]{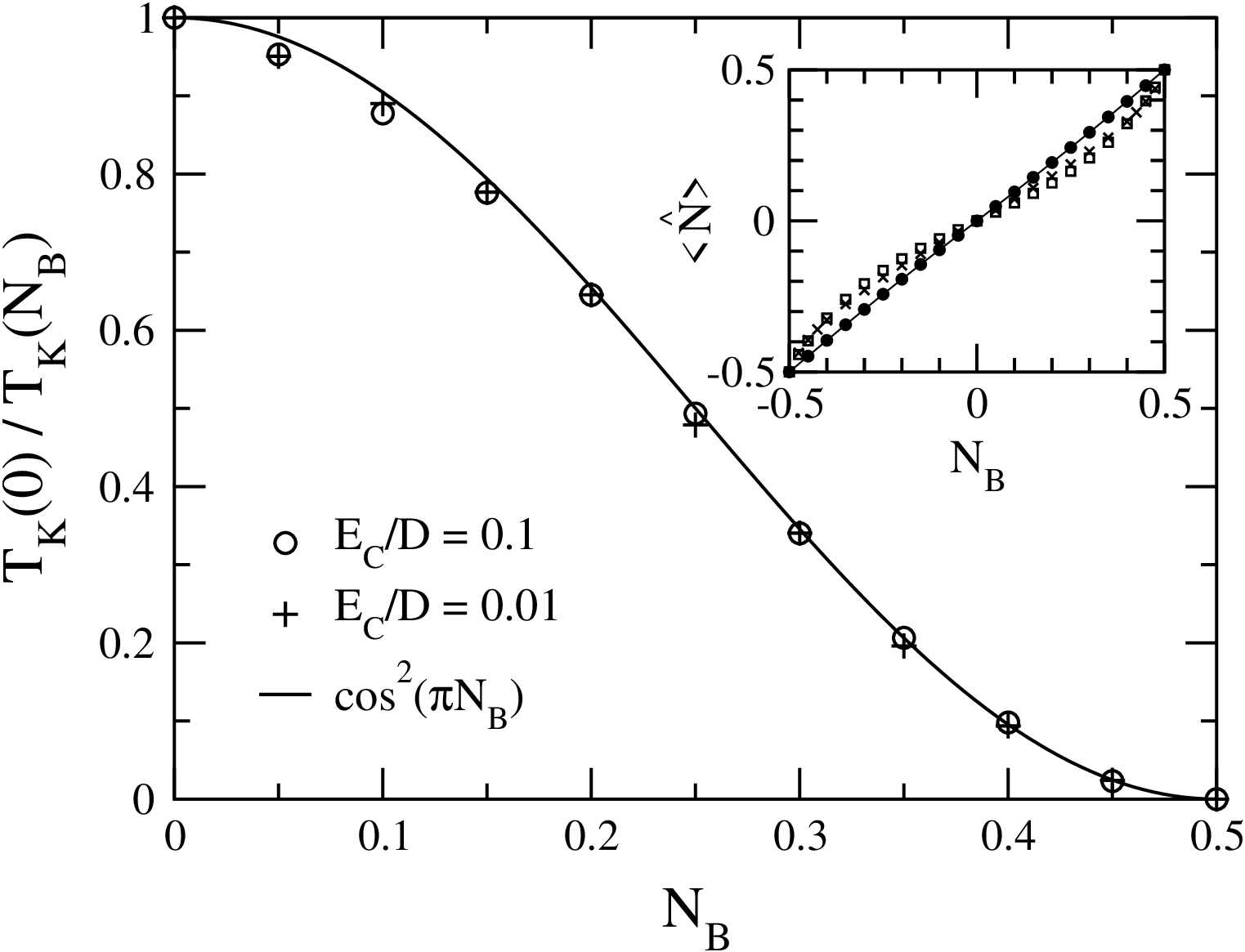}
}\vspace{-5pt}
\caption{$1/T_K$ vs $N_B$ for $U = \epsilon_d = 0$ and
         $\Gamma_L/D = 0.1$. Here $E_C/D$ equals $0.1$ (circles)
	 and $0.01$ (pluses). For each combined value of $N_B$
	 and $E_C$, the coupling $\Gamma_B$ is tuned to the
	 two-channel point $\Gamma_B^{\rm 2CK}$. As a function
	 of $N_B$, $\Gamma_B^{\rm 2CK}(N_B)$ varies by less
	 than $0.2\%$ for each fixed value of $E_C$. The ratio
	 $T_K(0)/T_K(N_B)$ is well described by $\cos^2(\pi N_B)$
	 (solid line).
         Inset: The box charge $\langle \hat{N} \rangle$ vs $N_B$
         for $\Gamma_L = E_C = 0.1 D$ and $U = \epsilon_d = 0$.
         Here $\Gamma_B/\Gamma_L$ equals $1$ (squares), $1.72$
         (filled circles + solid line), and $3.24$ (crosses).
         The Coulomb staircase is completely washed out for
         $\Gamma_B = \Gamma_B^{\rm 2CK}(0) = 1.72\Gamma_L$, but is
         gradually recovered upon departure from
	 $\Gamma_B^{\rm 2CK}(0)$.}
\label{fig:fig2}
\end{figure}

Equation~(\ref{chi_Matveev}) formally describes the limit
$E_C \ll \Gamma_L$, since $\Gamma_L$ serves as the effective
bandwidth for the 1D model used (see Ref.~\onlinecite{LSA03},
Sec.~III). Surprisingly, we find good agreement with the NRG
results for $U = 0$ even for $E_C$ as large as $\Gamma_L$.
(i) Scanning $N_B$ for fixed $E_C$ and $\Gamma_L$, we find a
    2CKE for all gate voltages, with a Kondo
    temperature that varies as $T_K(N_B) \propto 1/\cos^{2}(\pi N_B)$
    (Fig.~\ref{fig:fig2}).
(ii) Consistent with the notion of perfect transmission, only small
     variations are found in $\Gamma_B^{\rm 2CK}(N_B)$ as a function of
     $N_B$ (less than $0.2\%$), falling within our numerical percision.
(iii) The Coulomb staircase in the charging of the box is %% completely
      washed out for $\Gamma_B = \Gamma_B^{\rm 2CK}(0)$, but is
      gradually recovered as $\Gamma_B$ sufficiently departs from
      $\Gamma_B^{\rm 2CK}(0)$, whether above or below (inset of
      Fig.~\ref{fig:fig2}).
(iv) Upon reducing $E_C/\Gamma_L$ from $1$ to $0.081$,
     the ratio $\Gamma_B^{\rm 2CK}(0)/\Gamma_L$ steadily
     decreases from $1.72$ to $1.11$. This suggests the limit
     $\Gamma_B^{\rm 2CK} \to \Gamma_L$ as $E_C \to 0$, matching the
     condition for perfect transmission for two noninteracting leads.
Thus, the spin 2CKE for $U = 0$ is well described by
Eq.~(\ref{chi_Matveev}).

\begin{figure}
\centerline{
\includegraphics[width=75mm]{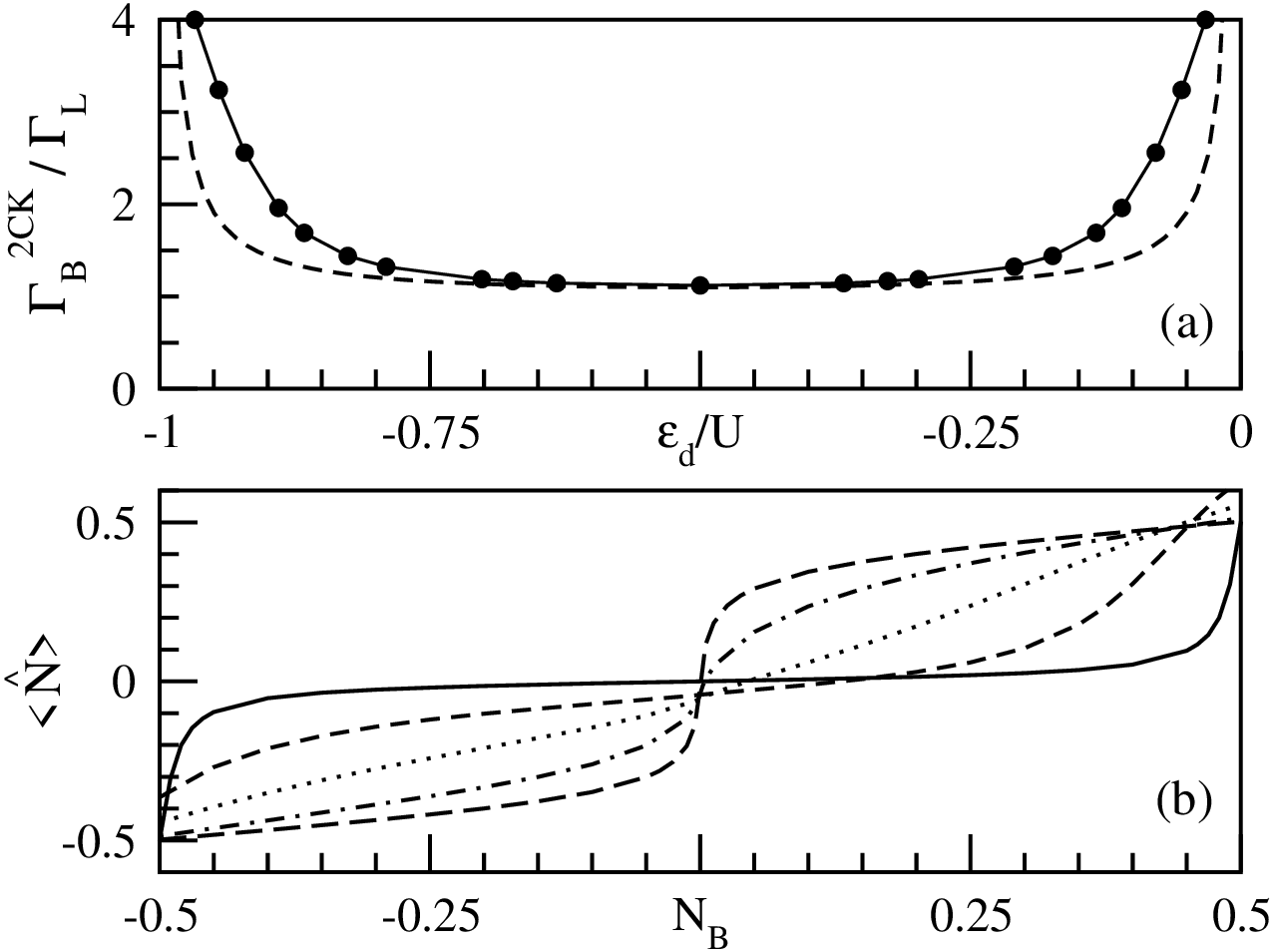}
}\vspace{-5pt}
\caption{(a) The two-channel line $\Gamma_B^{\rm 2CK}/\Gamma_L$
         vs $\epsilon_d$ for $U/D = 2$, $\Gamma_L = E_C = 0.1D$,
         and $N_B = 0$. Dashed line: The local-moment estimate of
         Ref.~\onlinecite{OGG03}.
         (b) The charge curve $\langle \hat{N} \rangle$ vs $N_B$,
         for $\epsilon_d/D = -1, -1.581, -1.733, -1.843$, and
         $-1.935$ (solid, dashed, dotted, dot-dashed, and
         long-dashed line, respectively). For each value of
         $\epsilon_d$, $\Gamma_B$ is tuned to the corresponding
         $N_B =0$ value of $\Gamma_B^{\rm 2CK}$, depicted in
         panel (a).}
\label{fig:fig3}
\end{figure}

\begin{figure}[b]
\centerline{
\includegraphics[width=75mm]{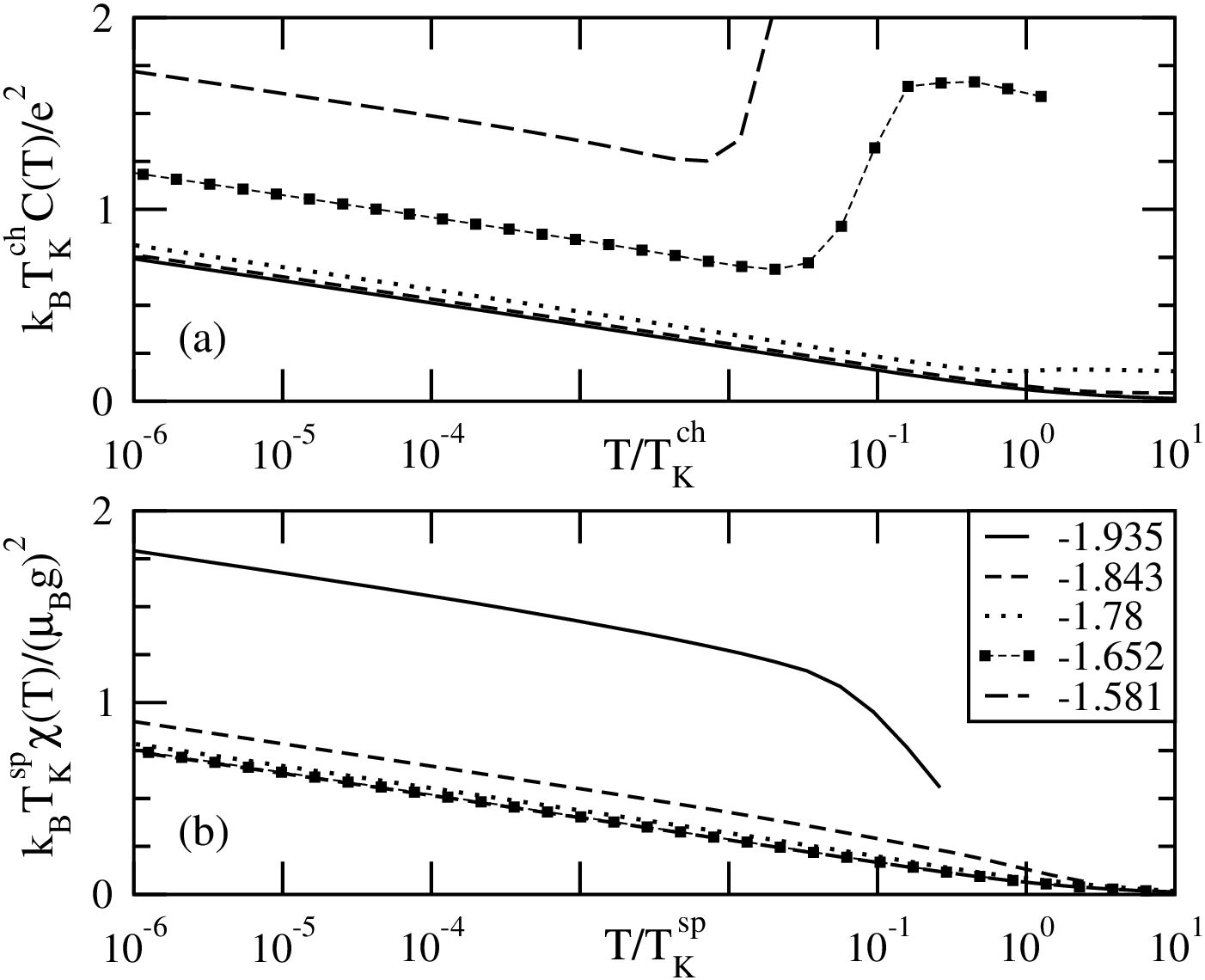}
}\vspace{-5pt}
\caption{(a) The capacitance of the box and (b) the magnetic
         susceptibility of the dot, for $U/D = 2$,
         $\Gamma_L = E_C = 0.1D$, $N_B = 0$, and different
         values of $\epsilon_d/D$, as specified in the legends.
         For each $\epsilon_d$, $\Gamma_B$ is tuned to the
         corresponding $N_B =0$ value of $\Gamma_B^{\rm 2CK}$.
         Both $\chi(T)$ and $C(T)$ diverge logarithmically
	 as $T \to 0$, but with different Kondo scales
         $T_K^{\rm sp}$ and $T_K^{\rm ch}$, extracted from
	 the slopes of their $\ln(T)$ diverging terms. In
         going from $\epsilon_d/D = -1.581$ to $\epsilon_d/D = -1.935$,
         $k_B T_K^{\rm sp}/D$ takes the values $0.0022, 0.0054,
         0.0457, 0.187$, and $1.72$, while $k_B T_K^{\rm ch}/D$
         equals $1.035, 0.36, 0.035, 0.009$, and $0.00138$,
         respectively.}
\label{fig:fig4}
\end{figure}

So far we have considered a symmetric dot, and varied
$U = - 2\epsilon_d$. In reality, however,
%% the Coulomb repulsion
$U$ is large and fixed. The experimentally tunable
parameters are the dot level $\epsilon_d$, the dimensionless
gate voltage $N_B$, and, to a lesser degree, the tunneling
rates $\Gamma_L$ and $\Gamma_B$. In Figs.~\ref{fig:fig3} and
\ref{fig:fig4} we explore the 2CKE as a function
$\epsilon_d$, for $U/D = 2$ and $\Gamma_L = E_C = 0.1D$.
Figure~\ref{fig:fig3}(a) shows the two-channel line
$\Gamma_B^{\rm 2CK}/\Gamma_L$ versus $\epsilon_d$, for $N_B = 0$.
The two-channel line separates two distinct Fermi liquids,
where the dot is coupled more strongly to the box (for $\Gamma_B$
above the line) or to the lead (below and to the side of the line).
Deep in the local-moment regime there is good agreement with the
estimate of Ref.~\onlinecite{OGG03} based on the Schrieffer-Wolff
transformation (dashed line). However, large deviations develop
as one approaches the mixed-valent regime.

%% For comparison, the analytic estimate of Ref.~\onlinecite{OGG03}
%% for $\Gamma_B^{\rm 2CK}/\Gamma_L$ is plotted by the dashed line.
%% There is good agreement deep in the local-moment regime, but
%% large deviations develop as one approaches the mixed-valent
%% regime.

Fixing $\Gamma_B$ at the $N_B =0$ value of $\Gamma_B^{\rm 2CK}$,
the shape of the charge step dramatically changes upon going from
the local-moment to the mixed-valent regime [see
Fig.~\ref{fig:fig3}(b)]. For $\epsilon_d = -U/2$, one recovers a
conventional Coulomb-blockade staircase, with charge plateaus at
integer units of charge. Upon decreasing $\epsilon_d$, the Coulomb
staircase is gradually smeared, until it is essentially washed out
for $\epsilon_d/D \approx -1.73$. Upon further decreasing $\epsilon_d$,
there is a reentrance of the Coulomb staircase. However, the degeneracy
points are shifted to integer values of $N_B$, and the charge plateaus
occur at half-integer units of charge.

To understand this surprising shift of the Coulomb staircase, we note
that it happens for $\Gamma_B^{\rm 2CK}$ several times larger than
$\Gamma_L$ and $E_C$. We therefore diagonalize first the local problem
of the dot and the two box degrees of freedom directly coupled to it,
before incorporating the smaller scales $\Gamma_L$ and $E_C$.
For $U \gg t_B \gg U + \epsilon_d$, the ground state of the local
problem is a strong admixture of the dot and box degrees of freedom,
whose respective occupancies are $3/2$ and $1/2$. Switching on $E_C$,
the remaining box degrees of freedom not directly coupled to the dot
thus experience an effective charging-energy interaction similar to
that of Eq.~(\ref{initial_Hamiltonian}), but with $N_B$ shifted
by half an integer. The resulting integer charge plateaus for the
remaining box degrees of freedom translate then to half-integer
charge plateaus for the entire box.

\begin{figure}[t]
\centerline{
\includegraphics[width=75mm]{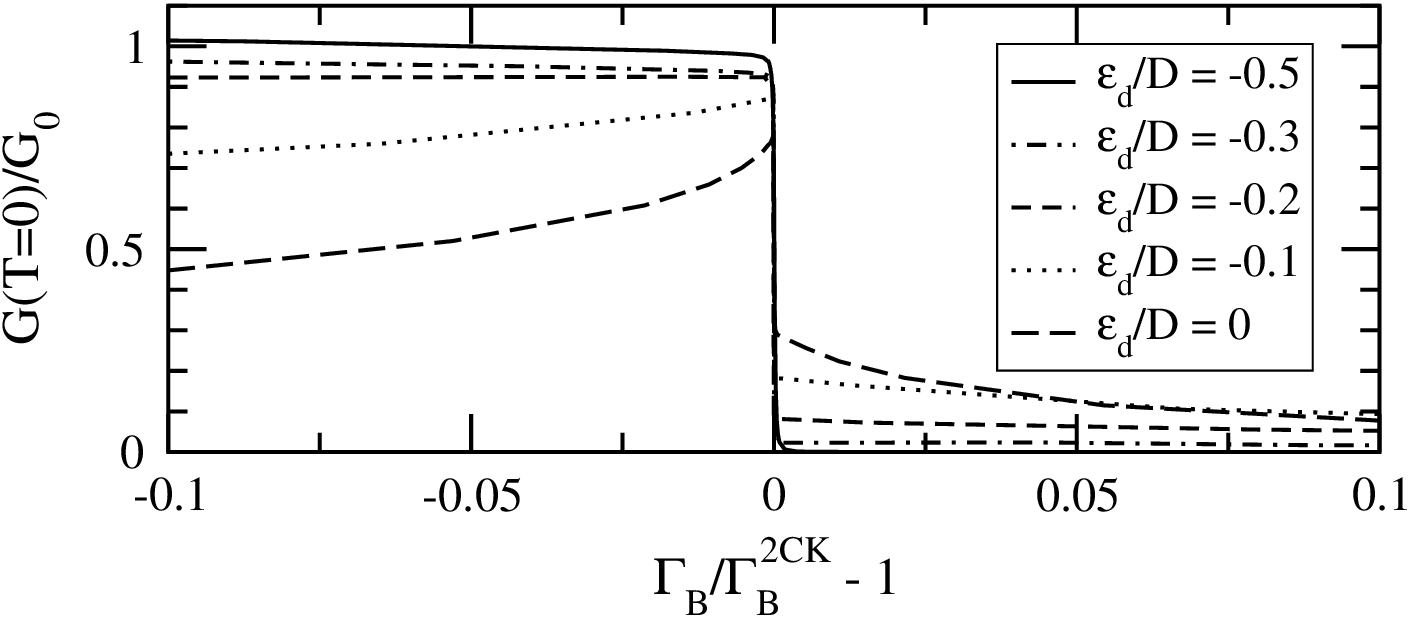}
}\vspace{-5pt}
\caption{The zero-temperature conductance of the two-lead device
         proposed in Ref.~\onlinecite{OGG03}, for $U/D = 1$,
         $E_C/D = 0.1$, $N_B = 0$, and different values of
         $\epsilon_d$. The tunneling rates $\Gamma_l$ and
         $\Gamma_r$ are kept fixed in all curves, with
	 $\Gamma_l + \Gamma_r = 0.1D$. Here $G_0 h/2e^2$ equals
	 $4\Gamma_l \Gamma_r/(\Gamma_l + \Gamma_r)^2$.}
\label{fig:fig5}
\end{figure}

The most striking feature of particle-hole asymmetry is the
entanglement of spin and charge within the 2CKE that develops.
As demonstrated in Fig.~\ref{fig:fig4} for $N_B = 0$ and
different values of $\epsilon_d \neq -U/2$, there is a
simultaneous $\ln(T)$ divergence of the dot susceptibility
$\chi(T)$ and the box capacitance
$C(T) = (e^2/2 E_C)d\langle \hat{N} \rangle /d N_B$, when
$\Gamma_B$ is tuned to the corresponding $N_B =0$ value of
$\Gamma_B^{\rm 2CK}$. Hence, the resulting 2CKE is neither
of pure spin nor of pure charge character, but rather
involves both sectors. Quite remarkably,
the degeneracy point where $C(T \to 0)$ diverges
is pinned at $N_B = 0$ for all $\epsilon_d$, although the
charge curves of Fig.~\ref{fig:fig3}(b) show no particular
symmetry about this point. Moreover, there are two distinct
Kondo scales, $T_K^{\rm sp}$ and $T_K^{\rm ch}$, extracted
from the slopes of the $\ln(T)$ diverging terms in $\chi(T)$
and $C(T)$. Upon decreasing $\epsilon_d$ from $-U/2$ to $-U$,
$T_K^{\rm sp}$ monotonically increases while $T_K^{\rm ch}$
monotonically decreases. This marks a continuous transition
from a predominantly spin 2CKE deep in the local-moment
regime, to a predominantly charge 2CKE, reminiscent of
Matveev's scenario,~\cite{LSA03,Matveev91} in the strongly
mixed-valent regime.

Experimentally, the relevant temperature scale is the crossover
temperature $T_0$, below which the 2CKE sets in. Estimating
$T_0$ from the NRG level flow, we find that it roughly traces
$T_{\rm min} = {\rm min}\{T_K^{\rm sp}, T_K^{\rm ch}\}$. The latter
scale is greatly enhanced when spin and charge are strongly
entangled, reaching a maximum of $k_B T_{\rm min} \sim 0.4E_C$
for the model parameters of Fig.~\ref{fig:fig4}. Hence, the
conditions for observing the 2CKE in realistic quantum-dot
devices are most favorable when the spin and charge degrees
of freedom are strongly entangled.

The dot susceptibility is very useful theoretically for analyzing
the 2CKE, but difficult to measure for a single dot. In
Fig.~\ref{fig:fig5} we depict the $T = 0$ conductance, $G(0)$,
for the two-lead device proposed in Ref.~\onlinecite{OGG03}. Here
the single lead is replaced with two separate leads, characterized
by the tunneling rates $\Gamma_l$ and $\Gamma_r$.
In equilibrium, this setting is equivalent~\cite{OGG03} to a single
lead with $\Gamma_L = \Gamma_l + \Gamma_r$. At $T = 0$, the
conductance is proportional to the dot spectral function at the
Fermi energy, which we calculate using the NRG. Fixing $\Gamma_l$
and $\Gamma_r$, $G(0)$ drops abruptly as $\Gamma_B$ crosses
$\Gamma_B^{\rm 2CK}$. However, in contrast to the treatment of
Ref.~\onlinecite{PBGvD03}, the height of the conductance step is not
fixed. It decreases in size with increasing charge fluctuations on
the dot. Indeed, away from the Kondo limit $G(0)$ neither vanishes
for $\Gamma_B > \Gamma_B^{\rm 2CK}$ nor reaches the unitary limit
for $\Gamma_B < \Gamma_B^{\rm 2CK}$. In the mixed-valent regime,
$G(0)$ actually develops a maximum at
$\Gamma_B = \Gamma_B^{\rm 2CK} + 0^{-}$, reflecting the crossover
to a Matveev-type charge 2CKE.

In summary, we found a line of two-channel fixed points in a
double-dot device, characterized by the number of excess
electrons in the box and by the ratio of the tunneling rates.
These fixed points govern the crossover from one Fermi liquid to
another, and are experimentally detectable by a sharp drop in the
conductance. Charge and spin degrees of freedom are generally
entangled, and decouple only at special particle-hole symmetry
points such as $N_B = 0$ and $\epsilon_d=-U/2$. The crossover
temperature to the two-channel Kondo effect is greatly enhanced
away from the local-moment regime, making this exotic effect
accessible in realistic quantum-dot devices. 

We have benefited from fruitful discussions with L. Glazman,
D. Goldhaber-Gordon, Y. Oreg, D. Orgad, and R. Potok.
E.L. and A.S. were supported in part by the Centers of
Excellence Program of the Israel Science Foundation.
F.B.A. acknowledges funding of the NIC, Forschungszentrum
J\"ulich, under Project No. HHB00.


\begin{thebibliography}{99}
\bibitem{comment_on_susceptibility} The dot susceptibility was
        computed from the dot magnetization for a small local
	magnetic field.
\bibitem{forthcoming} Details of the calculation will be
        published elsewhere.
\end{thebibliography}
\end{document}